# An SOA-Based Big Data Management Framework for Primary Healthcare Centers in Bahrain

**Hasan Abdulla [1], Fatema AlJazeeri [2], Shaikha Hasan [3], Zainab Almahafdha [4], Ehab Adwan [5]**

[1] habdulla@phc.gov.bh, Operations & Services, Primary Healthcare Centers, Manama, Kingdom of Bahrain
[2] fhahmed@uob.edu.bh, College of Information Technology, University of Bahrain, Sakhair, Kingdom of Bahrain
[3] saliahmed@uob.edu.bh, College of Information Technology, University of Bahrain, Sakhair, Kingdom of Bahrain
[4] z.almahafdha@bub.bh, College of Engineering & Technology, British University of Bahrain, Manama, Kingdom of Bahrain
[5] eadwan@uob.edu.bh, College of Information Technology, University of Bahrain, Sakhair, Kingdom of Bahrain

**Abstract**

The rapid data growth in the healthcare industry has presented significant challenges in managing and leveraging big data effectively. This research aims to design and apply a service-oriented architecture-based approach for managing big data for a Primary Healthcare Center in Bahrain (SOA-MHC) to overcome these challenges. The aim is to investigate the application of Service-Oriented Architecture (SOA) principles in enterprise architecture for managing big data in the healthcare center. The research employs the Design Science Research Methodology (DSRM) to guide the development of the SOA-MHC artifact. The six phases of the DSRM, including problem identification and motivation, defining objectives of a solution, design and development, demonstration, evaluation, and communication, are followed to ensure a comprehensive approach to the research.

The SOA-MHC uses a layered architecture comprising a Presentation, Service, and Data Integration Layer. The architecture leverages SOA principles such as loose coupling, reusability, and interoperability to create a modular and adaptable architecture capable of efficiently handling the complexities of healthcare big data. The research utilizes diverse data collection methods, including stakeholder interviews, document reviews, observations, and benchmarking research, to gather insights and inform the design and development of the SOA-MHC model.

The study's findings highlight the significant benefits of the SOA-MHC architecture, including improved data integration and interoperability, enhanced data analytics capabilities, operational efficiency gains, and improved patient engagement and satisfaction. The architecture demonstrates the effectiveness of the SOA-based approach in addressing the challenges of big data management in healthcare. The research contributes to the growing knowledge of applying SOA principles in healthcare and provides valuable insights for practitioners and researchers. The implementation of the SOA-MHC architecture provides an example for other healthcare institutions that want to utilize big data and service-oriented architectures to transform their data management methods and enhance patient care quality. The implementation of the SOA-MHC architecture shows an example for other healthcare institutions that want to use big data and service-oriented architectures to change their data management methods and enhance patient care quality.

## 1. Introduction

In the rapidly evolving healthcare sector, the application of Service-Oriented Architecture (SOA) to enterprise architecture (EA) is increasingly recognized as a valuable strategy for managing big data effectively [1]. SOA provides a structural approach where systems are comprised of interconnected, reusable services, offering the flexibility and interoperability necessary to leverage healthcare's vast and diverse data sources [2]. The healthcare industry's big data, derived from electronic medical records, imaging, genomics, and various digital platforms, presents significant opportunities to revolutionize patient care and optimize operations. However, the challenge of siloed datasets and fragmented data management systems often hinders the ability to gain integrated insights [3]. Implementing a SOA-based architecture can address these challenges by facilitating the orchestration of various data services into comprehensive applications, thus meeting the healthcare industry's demands for scalability, agility, and actionable analytics [4].

Moreover, SOA can contribute to developing agile and scalable platforms essential for real-time analytics and decision support in a healthcare context, enabling complex data integration into coherent and patient-centric care models [5]. As such, SOA-based EA frameworks are poised to play a crucial role in optimizing big data management in healthcare settings, providing a pathway to improved patient outcomes and more efficient healthcare delivery.

This paper examines how applying SOA principles facilitated the Primary Healthcare Center's (PHC) big data integration, analytics, and governance through an enterprise architecture based on loose coupling, reusability, and interoperability. It provides empirical insights into the real-world implementation of a SOA-powered data architecture within a healthcare organization. The critical research problem addressed in this paper is the challenge of effectively managing and leveraging big data in healthcare organizations to improve patient care and operational efficiency. Specifically, it focuses on how SOA-based enterprise architecture can facilitate data integration, analytics, and governance in primary healthcare.

The aim of this research is to apply SOA principles in enterprise architecture for managing big data in a primary healthcare center located in Bahrain. By exploring this real-world implementation, the study seeks to gain valuable insights into the effectiveness of SOA-based approaches in handling the complexities of healthcare data management. To achieve this aim, the research will focus on several key objectives. First, it will examine the challenges and opportunities associated with big data management in the healthcare sector, providing a comprehensive understanding of the current landscape and the potential benefits of effective data management strategies. Second, the study will explore the role of SOA-based enterprise architecture in addressing these challenges, highlighting how the principles of SOA can facilitate data integration, interoperability, and scalability.

Furthermore, the research will analyze the implementation of a SOA-powered data architecture in the Bahraini primary healthcare center, offering a detailed case study of the real-world application of these principles. Through this analysis, the study aims to identify the benefits, challenges, and lessons learned from the implementation process, providing valuable insights for other healthcare organizations considering similar approaches to big data management.

The research at hand is based on two strands, including healthcare data management and enterprise architecture. This research contributes in two ways. Firstly, it contributes to the field by providing empirical evidence on the effectiveness of SOA-based enterprise architecture for big data management in a healthcare context. The research demonstrates the practical application of SOA principles and their impact on data integration, analytics, and governance in a primary healthcare setting through a real-world case study. Moreover, the study offers valuable insights and lessons learned from implementing SOA principles, which can guide other healthcare institutes in considering the adoption of SOA-based approaches to big data management. Secondly, this research contributes to understanding how SOA can be effectively utilized to manage big data in healthcare organizations. Ultimately, this research promotes the adoption of SOA-based enterprise architecture in healthcare organizations, as it can create unified, flexible big data ecosystems that facilitate the integration and analysis of vast amounts of healthcare data, potentially improving evidence-based care and patient outcomes.

The rest of the paper is structured as follows: Section 2 presents a literature review on EA in healthcare and the application of SOA in big data management. Section 3 describes the proposed methods, including data collection and analysis techniques, and introduces the SOA-MHC, its key components, and its relevance to big data management in healthcare. Section 4 discusses the case study's findings, highlighting the challenges faced, the benefits realized, and the implications of the findings for healthcare organizations and the broader field of SOA. Finally, Section 5 concludes the paper by summarizing the key takeaways and identifying potential areas for future research.

## 2. Systematic Literature Review (SLR)

Big data in healthcare, or the large amount of health data, can change how healthcare is provided, make patient outcomes better, and drive innovation in the sector [6]. The deluge of data from various digital healthcare technologies necessitates a robust and strategic management framework. EA has proven critical for aligning an organization's processes, systems, and technology infrastructure with overarching strategic goals [7]. In healthcare, EA offers a disciplined methodology for managing big data, which is essential for maintaining data quality, security, interoperability, and informed decision-making. This section aims to present a literature review on the integration of big data in healthcare, highlighting its sources, opportunities, and challenges. Moreover, it will review the benefits of adopting EA in healthcare and the common EA frameworks used in healthcare.

### 2.1 Literature Profiling Methodology

This section details the analysis methodology utilized to investigate the current state of the art about the study objectives. The first step was identifying key search terms based on the main research question. These key terms are shown in Table 1. Next, the search query was refined by adding multiple keywords using Boolean operators (AND, OR) to yield more targeted results. After an initial search, the results were filtered by skim reading to eliminate irrelevant studies and identify high-quality literature aligned with the research aims. This rigorous screening process was applied when searching primary and secondary data sources to ensure the selection of top-tier, relevant journals, and articles, emphasizing sources indexed in Scopus with high-impact factors. References were carefully evaluated for relevance to the research questions and academic rigor.

**Table 1** List of search keywords used to identify relevant literature for big data in healthcare.

| Searching Keywords | | |
|---|---|---|
| Big Data | Healthcare | Clinical Decision Support |
| Analytics | Health | Method |
| Adoption | EHR | Model |
| Framework | EMR | Clinic |
| SOA | Service-Oriented | Architecture |

## 2.2 Big Data in Healthcare

The emergence of big data in healthcare transforms patient information management, signifying a paradigm shift in data collection, storage, analysis, and use [8]. Big data in healthcare encompasses a vast array of structured and unstructured data derived from electronic health records, medical imaging, wearable devices, genetic sequencing, social media, and other sources [9]. The exponential growth in such data—characterized by volume, velocity, and variety—presents formidable challenges and substantial opportunities for healthcare organizations to improve patient care, propel research, and enhance operational effectiveness [10].

Understanding the profound impact of big data in healthcare necessitates examining its defining attributes and the diverse sources fueling its expansion [11]. The following discussion will delve into the essential characteristics of big data relevant to healthcare, the varied origins of healthcare data, and the implications of competently managing and analyzing such extensive datasets. Through this exploration, insight can be gained into big data's transformative potential and complexities in the modern healthcare landscape.

### *2.2.1 Definition and characteristics of big data in healthcare*

Big data in healthcare has created huge amounts of information that can be used to improve the health and well-being of patients and the effectiveness and productivity of operations [12]. A defining feature of healthcare big data is the sheer volume, as exemplified by the vast quantities of data coming from electronic health records (EHRs) and medical imaging modalities [13]. Additionally, the velocity at which healthcare data accumulates is notable, especially considering the real-time data generated through Internet of Things (IoT) devices and telemedicine platforms [14]. The extensive volume and rapid velocity of healthcare data signify the big data paradigm shift within the sector. However, to fully realize the benefits of big data in transforming healthcare delivery and administration, robust analytics architectures must be implemented to manage the scale and complexity of healthcare data while upholding security, privacy, and regulatory compliance.

Furthermore, the variety of data, ranging from structured formats like EHRs to unstructured data from clinical notes and images, adds to the complexity of data management in the healthcare sector [15]. Veracity, or the trustworthiness of the data, is crucial, as the quality of healthcare decisions is only as good as the data, they are based on [16]. Ultimately, the value derived from big data in healthcare lies in its ability to provide actionable insights for improved clinical decision-making and patient care [17].

To harness big data's full potential, healthcare organizations face challenges such as ensuring data privacy and security, achieving interoperability among disparate health information systems, and developing a skilled workforce capable of analyzing and interpreting complex datasets. Despite these challenges, big data's opportunities in advancing healthcare are significant, underscoring the need for robust enterprise architecture frameworks that can support the holistic strategic alignment of complex healthcare data.

### *2.2.2 Sources of big data in healthcare*

The healthcare industry has been a significant catalyst for the proliferation of big data, with numerous sources producing vast quantities of information. At the vanguard are EHRs, which furnish comprehensive digital documentation of patient data that is fundamental for enhancing healthcare delivery and enabling clinical research [18]. Medical imaging technologies also significantly contribute to expanding healthcare data, offering vital diagnostic and treatment insights derived from high-resolution images across modalities such as X-ray, MRI, and CT scans [19]. Furthermore, genomic, and molecular data from next-generation sequencing, proteomics, and other omics technologies are amassing at unprecedented rates, harboring the potential to transform precision medicine through personalized therapies. Additional sources fueling big healthcare data include real-time telemetry from wearable devices, medical IoT sensors, information from health insurance claims, and more [2], [20]. The deluge of data from these multifaceted sources underscores the disruptive nature of big data across the healthcare ecosystem while necessitating robust analytics capabilities to harness its latent value.

The advent of wearable devices and the IoT has unleashed a new wave of patient-generated health data, enabling continuous monitoring and personalized care [21]. Genetic and genomic data further enhance the capability for precision medicine, with large-scale genomic datasets underpinning targeted therapy development [22]. Insurance claims and administrative data provide a window into healthcare utilization and resource management, which is crucial for identifying health trends and high-risk patient groups. Social media and patient-reported outcomes enrich these data sources with firsthand patient experiences, offering a unique perspective often absent from clinical data [13].

Clinical trials and biomedical research datasets constitute indispensable sources of evidence for evaluating novel medical interventions and propelling health science discoveries [22]. However, the integration and harmonization of these heterogeneous, fragmented data silos presents a formidable challenge. Adopting robust data management strategies and strict conformance to interoperability standards are imperative [14]. Specifically, standardized clinical trial data models, such as the Clinical Data Interchange Standards Consortium (CDISC) standards, can significantly enhance research data's interoperability, exchange, and reuse. Additionally, applying ontologies and semantic technologies can enable richer integration and analysis across diverse datasets [23]. Furthermore, continued optimization of clinical research and trial data curation, documentation, sharing, and governance processes is needed [24]. Addressing these multifaceted data integration and interoperability challenges will be instrumental in unleashing the utility of clinical research data to inform evidence-based care and fuel biomedical insights.

#### *2.2.3 Challenges and opportunities associated with big data in healthcare.*

Integrating and analyzing big data within healthcare systems poses significant challenges and opportunities for improving patient care and operational efficiency. One of the primary challenges is achieving interoperability among diverse healthcare information systems, which is essential for the seamless exchange and utilization of health-related data [18], [21]. Moreover, maintaining the quality and accuracy of the data is critical, given that healthcare decisions rely on precise and reliable information [25].

Alongside the opportunities presented by healthcare big data, significant challenges exist regarding privacy, security, and regulatory compliance. Healthcare data frequently encompasses susceptible personal information, necessitating stringent safeguards to protect patient privacy and prevent improper data usage or breaches [26]. Rigorous adherence to regulations, including the Health Insurance Portability and Accountability Act (HIPAA) and General Data Protection Regulation (GDPR), is imperative when handling healthcare data [27]. Additionally, the effective harnessing of healthcare big data is constrained by a shortage of professionals possessing competencies in pertinent domains like data science, healthcare informatics, clinical contexts, and regulatory policies [21], [28]. Developing specialized educational programs to cultivate such cross-disciplinary expertise and instituting governance frameworks that embed privacy and ethics considerations into data analytics initiatives will be vital to responsibly unleashing the power of healthcare big data. Patient trust and stakeholder buy-in are contingent upon transparent data governance and the assurance that privacy and security remain top priorities.

The advent of big data analytics in healthcare introduces salient opportunities to advance patient-centered care and improve population health outcomes. Most notably, big data-driven approaches can enable personalized or precision medicine, allowing preventative interventions and treatments to be more precisely tailored to an individual's genomic profile, lifestyle factors, and health status [29]-[30]. Additionally, predictive analytics leveraging patient data aggregated across populations can prove invaluable for early disease detection and timely intervention [31]. At the population level, big data-derived insights can strengthen population health management initiatives and empower public health authorities to make data-driven decisions regarding resource allocation, outbreak prevention, and community health priorities [32]. However, realizing this value potential necessitates building robust and ethical data analytics capabilities spanning data integration, predictive modeling, and transparent governance. Overall, healthcare big data heralds' immense possibilities to usher in an era of democratized, personalized, and evidence-based medicine. However, it must be harnessed judiciously and equitably to maximize benefits for all patient groups.

Operational efficiencies and cost reductions are also achievable through the strategic analysis of big data, which can identify inefficiencies and optimize healthcare delivery [25]. Lastly, big data is a boon for healthcare research, offering a rich resource for innovation and developing new treatments and drugs [29]. To capitalize on these opportunities, healthcare organizations must navigate the complexities of data integration and invest in infrastructure and skills development to effectively manage and analyze large datasets [21]. By addressing these challenges, the healthcare sector can unlock the transformative potential of big data to improve patient care and operational efficiency.

### 2.3 EA in Healthcare

EA has emerged as a crucial approach for healthcare organizations to effectively manage their complex IT systems and align them with strategic objectives. As healthcare institutions face the challenges of rapidly evolving technologies, increasing data volumes, and the need for interoperability, adopting EA principles has become essential. EA provides a holistic view of an organization's IT landscape, enabling better decision-making, resource allocation, and process optimization. The following subsections delve into the definition and importance of EA in healthcare, the commonly used EA frameworks, and the benefits of adopting EA in healthcare organizations.

#### *2.3.1 Definition and importance of EA in healthcare*

In the healthcare sector, EA refers to the entire framework and process used to align the organizations' IT infrastructure with its strategic goals and objectives [33]. Within a healthcare institution, EA entails the strategic planning, design, and administration of IT systems, data, procedures, and technology. Utilizing EA in healthcare assists in ensuring interoperability between disparate systems, streamlining IT systems and processes, cutting maintenance costs through suitable planning, implementing robust security controls, and providing flexible and scalable IT solutions adaptable with patients' needs [21].

#### *2.3.2 EA frameworks used in healthcare (e.g., TOGAF, Zachman, FEA)*

The TOGAF framework has explicitly assisted healthcare organizations in navigating big data complexities by providing a structured approach tailored to the sector's unique challenges [34]. Leveraging big data in healthcare depends on the capability to manage and derive actionable insights from vast datasets [35]. Through its Architecture Development Method (ADM), TOGAF offers healthcare organizations a blueprint to develop and govern an EA that supports big data initiatives [36]. Applying TOGAF principles, healthcare organizations can improve architecture management, leading to enhanced data-driven strategies critical for patient care and efficiency [37]. As healthcare organizations address the multifaceted landscape of big data, TOGAF serves as an indispensable tool to organize and manage enterprise architectures in support of strategic objectives, streamlined processes, and ultimately improved patient care [6]-[7], [34]-[37].

#### *2.3.3 Benefits of adopting EA in healthcare organizations*

The utilization of information and digital health technology in healthcare organizations is experiencing a significant revolution. To help with this shift, EA has been gradually embraced by several healthcare organizations across the world [18]. Organizations gain various advantages from implementing EA [38]. EA may improve the efficiency of an organization's resources and skills in the field of digital health [21]. Moreover, Enterprise architecture makes it easier to find, assess, and implement appropriate security measures that are suited to the unique needs of healthcare organizations, especially considering the risks posed by cyberspace. [16] Furthermore, it is necessary to emphasize the significance of managing and protecting patient data [39]. Additionally, EA assists healthcare organizations in identifying and eliminating redundant processes, and thus streamlining the processes.

### 2.4 SOA for Big Data Management

SOA represents a strategic design approach that structures IT infrastructure as a collection of loosely coupled, interoperable services intended for reuse across diverse business processes and applications. Within big data management, SOA can prove instrumental in addressing the challenges of volume, variety, velocity, and veracity, thereby enhancing the value derived from data assets [40]. By applying SOA principles, organizations can construct a flexible architecture that integrates complex data systems and rapidly assembles services to accommodate evolving data needs [41]. This SOA-based approach to big data management facilitates scalability, adaptability, and alignment of IT services with business objectives, ensuring the data architecture supports critical operational requirements [42]. SOA provides a strategic paradigm for structuring big data architectures to tackle the four Vs, maximize data value, and sustain alignment with business priorities through reusable, interoperable services [40]-[42].

*2.4.1 Applicability of SOA principles to big data management*
The principles of SOA, including loose coupling, service reusability, and interoperability, are critical for addressing the complexities inherent in big data management [40]. Loose coupling facilitates the development of platform- and technology-agnostic data services, conferring the flexibility imperative for adapting to the rapid transformations' commonplace in big data ecosystems [42]. Service reusability, a cornerstone of SOA, enables organizations to repurpose existing data services across diverse applications and processes, potentially decreasing redundancy and improving efficiency [41]. Interoperability, essential given the heterogeneity of data sources and systems in big data settings, ensures data services can communicate and function cohesively despite variances in underlying technologies or formats [43]. These foundational SOA principles establish a strategic framework to address the volume, velocity, variety, and veracity challenges intrinsic to big data management [40]-[43].

*2.4.2 Integration of big data in SOA-based EA*
Integrating big data into a SOA requires the design of data services that can effectively handle the distinctive features of big data, which include large volume, high velocity, wide variety, and robust veracity. This integration often necessitates implementing services capable of processing and analyzing extensive volumes of structured and unstructured data in real-time or near-real-time. Leveraging technologies such as Hadoop and Spark are a common practice to achieve computational scalability and efficiency. Moreover, the integration process may include creating services that amalgamate and standardize data from various sources, using data virtualization or federation to foster a unified data ecosystem. [40]

Encapsulating these capabilities within reusable services enables a modular, agile data architecture to accommodate evolving business needs and data landscapes [41]. This SOA-based approach adapts to shifting imperatives and optimizes data asset utilization, driving innovation and strategic objectives. However, effective implementation necessitates surmounting integration complexities and upholding security, latency, quality, and governance standards within heterogeneous big-data environments. [41]

*2.4.3 SOA-based EA for big data management in various industries*
The utility of SOA principles for managing big data has been demonstrated extensively within the healthcare domain. For instance, integrating and harmonizing patient health records across disparate HER systems leveraging SOA facilitates real-time analytics and advances personalized medicine [11]. Additionally, the design of an enterprise financial data-sharing service center employing big data technologies illustrates SOA's potential for constructing flexible, scalable data architectures adept at processing high volumes of structured and unstructured data [42].

Furthermore, big data frameworks like Hadoop have been implemented within healthcare settings using SOA principles to govern healthcare data more effectively, providing a model for other sectors seeking to develop value-driven data architectures [44]. While the references do not directly support the retail, finance, and telecommunications examples, these use cases underscore how SOA creates adaptable, performant data architectures across industries wrestling with big data complexities, as substantiated by healthcare implementations. Further research into SOA-big data integration across domains would enrich the discourse.

**2.5 EA and Big Data Management in the Healthcare Context**
The healthcare industry is increasingly adopting big data technologies, including advanced analytics, to enhance patient care delivery, optimize operations, and catalyze research and innovation. However, effective management and utilization of healthcare big data present distinct challenges, such as ensuring data quality, security, and integration across the heterogeneous systems comprising modern healthcare ecosystems [45]. These hurdles require a robust, holistic EA approach that governs intricate, large-scale healthcare data environments [46].

Recent literature underscores the need to implement advanced data analytics within healthcare, citing significant potential to improve patient outcomes and operational efficiency [2]. Additionally, emerging research indicates growing interest in cloud-based architectures to enable scalable and flexible healthcare big data analytics [47]. However, gaps persist regarding real-world empirical assessments of EA implementations in healthcare settings. While potential benefits have been posited, further investigation is imperative to elucidate best practices and quantify the impact of EA adoption on healthcare delivery [48]. Future research should emphasize developing EA models tailored to healthcare contexts and evaluating their efficacy via rigorous case studies. Continued exploration of EA design and validation can inform data-driven, patient-centric healthcare transformation.

*2.5.1 Specific challenges and requirements for big data management in healthcare*

Managing big data presents multifaceted challenges for healthcare organizations that necessitate tailored EA approaches accounting for sector-specific contexts. Foremost, safeguarding patient privacy and data security is non-negotiable, requiring stringent adherence to regulations, including The Health Insurance Portability and Accountability Act (HIPAA) and General Data Protection Regulation (GDPR) compliance. [5] EA frameworks must integrate robust security controls to prevent unauthorized data access or breaches [46]. Furthermore, the heterogeneity of healthcare data from sources like electronic health records, devices, and wearables poses interoperability and data quality challenges. Standardized data models and governance are critical for effective integration and exchange [49].

Additionally, real-time analytics for clinical decision support demand high-performance, scalable architectures [33]. Finally, aligning initiatives with clinical and operational needs mandates IT and healthcare professionals' collaboration. EA frameworks must bridge this gap to ensure big data analytics responsively inform healthcare delivery [48]. Overall, nuanced EA approaches addressing these multifaceted sector-specific challenges are imperative for successful healthcare big data implementation.

*2.5.2 Existing research on the adoption of EA for big data management in healthcare organizations*

The application of EA frameworks to manage big data is garnering increasing attention within healthcare contexts. Cuzzocrea and Soufargi [46] proposed a cloud-based EA model to enable scalable healthcare big data analytics, demonstrating the potential of standardized EA approaches like TOGAF to address sector-specific complexities. However, Mohammed and Naaz [48] revealed obstacles in practical EA adoption, including deficient expertise, resources, and unfavorable organizational culture.

Therefore, empirical assessments of EA implementation challenges in healthcare warrant further exploration. Guraya [33] also advocated for a patient-centric EA approach fusing big data analytics to progress value-based care delivery. While EA holds promise for streamlining healthcare big data management, additional research is imperative to inform optimal integration strategies aligned with patient needs and organizational capabilities. The discourse would benefit from investigations quantifying the impacts and validating best practices for EA-enabled big data governance in real-world healthcare settings.

*2.5.3 Gaps in the literature and the need for further research on SOA-based EA for big data management in healthcare*

While the potential of SOA for big data management in healthcare shows promise, empirical assessments of SOA-based EA implementations in healthcare contexts still need to be explored, obstructing evaluations of real-world efficacy [50]. Additionally, detailed design and implementation guidance tailored to healthcare's specialized regulatory, privacy, and data complexity requirements must be improved [5]. Investigating organizational and cultural dynamics influencing SOA adoption would provide valuable insights, as these factors are critical in integrating new architectures into healthcare environments [4].

Furthermore, research examining the integration of SOA-based EA with emerging technologies like AI, blockchain, and precision medicine is limited but warranted to support comprehensive data strategies [51]. Advancing this scholarly discourse necessitates an interdisciplinary approach combining academic, healthcare, and technical perspectives [48]. In summary, while SOA shows promise for healthcare big data applications, additional empirical assessments, practical implementation frameworks, and explorations of supporting technologies enrich an understanding of how SOA-based EA can optimize healthcare delivery.

**Table 2** Recent Studies on Enterprise Architecture (EA) and Big Data in Healthcare

| Source | Year | Focus | Methodology | Key Findings |
|---|---|---|---|---|
| **[25]** | 2019 | Health referral enterprise architecture design in Indonesia | TOGAF framework | Proposed EA design for health referral system to improve healthcare services |
| **[16]** | 2019 | Framework of antecedents and benefits of EA in healthcare | Literature review | Developed framework linking EA antecedents and benefits in healthcare, emphasized need for stakeholder involvement |
| **[18]** | 2020 | Adoption of EA for healthcare in AeHIN member countries | Survey and interviews | Identified challenges and benefits of EA adoption in healthcare, emphasized need for collaboration and standardization |
| **[21]** | 2020 | EA for healthcare information exchange (HIE) cloud migration | Literature review and case study | Proposed EA framework for HIE cloud migration to improve interoperability and data sharing |

| Source | Year | Focus | Methodology | Key Findings |
|---|---|---|---|---|
| **[35]** | 2020 | Big data analytics for improved healthcare service and management | Literature review | Discussed applications, challenges, and future directions of big data analytics in healthcare |
| **[52]** | 2020 | Designing EA for public health center based on TOGAF ADM | TOGAF Architecture Development Method (ADM) | Developed EA blueprint for public health center to support decision making and improve services |
| **[31]** | 2021 | Adaptive EA design for digital healthcare platform | Case study and interviews | Developed adaptive EA design for digital healthcare platform to support industry 4.0 and society 5.0 initiatives |
| **[32]** | 2022 | Using EA to identify benefits of building information model data in healthcare | Case study | Demonstrated how EA can be used to identify benefits of BIM data in healthcare operations |
| **[34]** | 2022 | Analysis of big data architecture for healthcare service | Case study of public hospital | Analyzed big data architecture components and processes for healthcare, proposed reference architecture |

## 3. Proposed Method

The Design Science Research Methodology (DSRM) is a problem-solving paradigm advocated by [53]-[55] as a popular approach in Information Systems (IS) research for the development of IT artifacts, such as SOA-MHC. The critical research problem addressed in this paper is the challenge of effectively managing and leveraging big data in healthcare organizations to improve patient care and operational efficiency, specifically focusing on how SOA-based enterprise architecture can facilitate data integration, analytics, and governance in primary healthcare. DSRM is embraced to solve certain organizational problems and is a popular facilitator for the collaboration of technology, enterprises, and individuals in real-world scenarios. Due to the combination of qualitative and quantitative nature of the research at hand, with respect to the development of the SOA-MHC artifact, as advocated by [53]-[56], and its integration with enterprises, business processes, technology, and individuals, DSRM is considered the most suitable research design. Figure 1 illustrates the six phases of the DSRM based on the insights of [53]-[54], while the following subsections describe each phase in details.

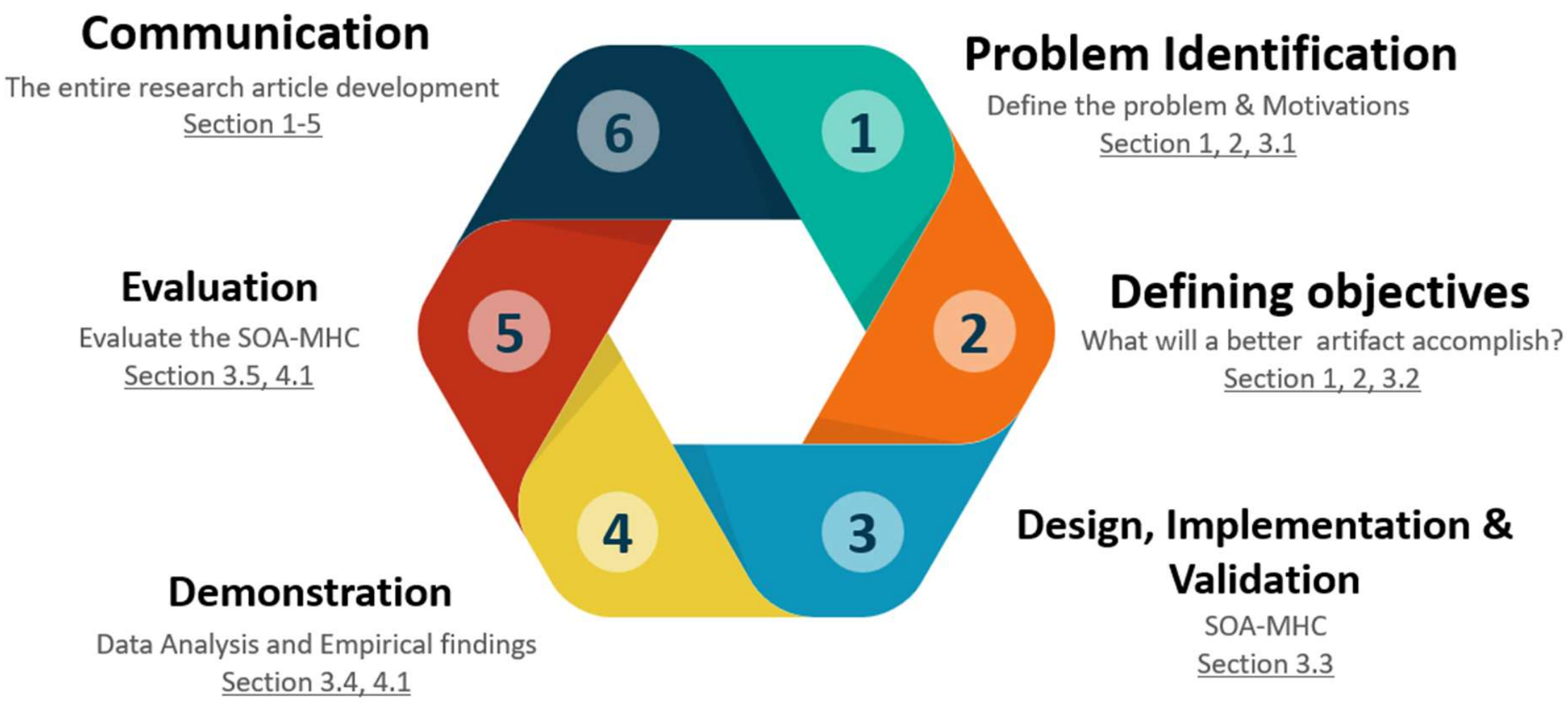

**Figure 1** The Design Science Research Methodology (DSRM) process model.

### 3.1 Problem identification & motivation

The healthcare industry has witnessed exponential growth in the volume, variety, and velocity of data generated from various sources, including EHRs. This explosion of healthcare data, often called "big data," presents challenges and opportunities for healthcare organizations. The primary motivation behind this research is to address the critical problems healthcare providers face in effectively managing and leveraging big data to improve patient care and operational efficiency.

One of the significant challenges in healthcare big data management is the need for interoperability and data integration across disparate systems [57]. Healthcare organizations often rely on multiple, siloed data sources that are not seamlessly connected, leading to difficulties in accessing and analyzing data holistically [58]. This data fragmentation hinders the ability to gain comprehensive insights into patient health, treatment outcomes, and operational performance [59]. Another significant problem is traditional data management systems' limited scalability and flexibility in handling the ever-growing volume and complexity of healthcare data. As data surges, healthcare organizations need help to efficiently store, process, and analyze large datasets in real time. This lack of scalability can result in delayed decision-making, reduced agility in changing healthcare needs, and missed opportunities for data-driven innovations [20].

Furthermore, ensuring data governance, security, and privacy is critical in healthcare big data management [20]. Healthcare data often contains sensitive patient information that must be protected in compliance with stringent regulations such as HIPAA and GDPR. Inadequate data governance and security measures can lead to data breaches, unauthorized access, and misuse of patient information, eroding trust and compromising patient privacy[21], [26]. The motivation for this research stems from the need to address these pressing challenges and unlock the potential of big data in healthcare. By developing SOA-MHC, this study aims to provide a scalable, flexible, and secure solution for integrating and analyzing healthcare data from diverse sources.

The SOA-MHC architecture leverages SOA principles like loose coupling, reusability, and interoperability to create a modular and adaptable system that efficiently handles healthcare big data complexities. It empowers healthcare professionals through seamless data integration, real-time analytics, and robust data governance, enabling data-driven decisions to improve patient outcomes and optimize resource use. This research is motivated by the potential impact of a SOA-based architecture on healthcare big data management in Bahrain and beyond. By demonstrating the SOA-MHC architecture's effectiveness in a real-world primary care setting, this study provides valuable insights and best practices that other healthcare organizations facing similar challenges can adopt. The problem identification and motivation address critical data integration, scalability, security, and governance issues in healthcare big data management. By developing a SOA-based solution, this study enables healthcare organizations to fully harness big data's potential, improve patient care, and drive industry innovation. Successful SOA-MHC implementation can transform how providers manage and use data, ultimately improving health outcomes and operational efficiency.

### 3.2 Defining objectives of a solution.

This research focuses on creating a SOA for a big data management tailored to a Bahraini primary healthcare center. The goal is to tackle the existing challenges associated with managing large volumes of healthcare data. The research has set several objectives to ensure a comprehensive approach to the problem. Firstly, it aims to integrate diverse data sources, such as electronic health records, by employing an interoperable architecture. Additionally, the research targets the design of a scalable and flexible architecture capable of handling the continuously increasing amounts of healthcare data.

A pivotal aspect of the research is enabling real-time analytics, crucial for making timely clinical decisions and providing personalized care. The research also emphasizes the importance of implementing robust data governance protocols to ensure the data's quality, privacy, and security. Promoting reusable and interoperable data services is critical to this objective, as well as optimizing performance to process data at a large scale efficiently. To make the system accessible to its users, developing a user-friendly interface is also a priority, facilitating easy access to data and its visualization.

The anticipated outcome is that by meeting these objectives, the SOA solution will empower the Healthcare Centre to unlock the full potential of its data, thereby improving patient outcomes and operational efficiency. The success of this research will be evaluated by how well it meets its objectives and its ability to demonstrate improvements in data integration, analytics capabilities, governance, and the overall quality of services. The insights garnered from this research are expected to benefit other healthcare organizations, considering the adoption of similar big data management strategies.

### 3.3 Design, Implementation & Validation

The design, implementation, and validation of the SOA-MHC architecture for the Bahraini primary healthcare center involved a comprehensive approach to ensure the effectiveness and suitability of the proposed solution. The process began with a thorough analysis of the healthcare center's existing data management practices, pain points, and requirements, which informed the design of the SOA-MHC. The design phase focused on creating a modular and scalable architecture that could effectively integrate and manage the healthcare center's big data while providing enhanced analytics capabilities. The implementation phase involved the development of the SOA-MHC components, including the Data Integration Layer, Service Layer, and Presentation Layer, along with integrating security and governance measures. Finally, the validation phase aimed to assess the implemented SOA-MHC model's performance, usability, and impact through rigorous testing, user feedback, and evaluation against predefined success criteria. The following sections delve into the details of each phase, highlighting the key considerations and outcomes.

*3.3.1 Data Collection*

A comprehensive data collection process was undertaken to understand the case thoroughly and construct the SOA-MHC. Data collection aimed to gather insights into the healthcare center's current data management practices, identify pain points and requirements, and inform the design and development of the SOA-MHC model. The Delphi technique was used to engage key stakeholders from the primary healthcare center, facilitating a deep understanding of the organization's data management challenges, needs, and expectations. These stakeholders included healthcare professionals, IT, academics, and EA experts.

Relevant documents and artifacts related to the healthcare center's data management practices and IT systems were reviewed. These documents included system documentation, data policies and procedures, and reports and analytics. The review helped to identify the current data practices and areas requiring alignment with the SOA-MHC model. Observation and shadowing sessions within the healthcare center were carried out to gain firsthand understanding of the data workflows and user interactions. Clinical workflows, data entry and processing, and data retrieval and analysis were observed. Documentation of the observations was done through field notes for further analysis.

To inform the architecture of the SOA-MHC, benchmarking and best practices research were reviewed. This involved reviewing academic literature, industry reports, and case studies related to SOA-based big data management in healthcare, studying the data management practices and architectures of peer healthcare institutions, and reviewing relevant industry standards and guidelines.

By employing these diverse data collection methods, a rich set of qualitative and quantitative data was gathered, which informed the understanding of the case and guided the construction of the SOA-MHC architecture. Insights gained from Delphi technique, document reviews, observations, user feedback, and benchmarking research formed the foundation for designing a comprehensive and tailored solution that addresses the specific needs and challenges of the Bahraini primary healthcare center.

*3.3.2 Data Validation*

To ensure the validity and reliability of the data collected for constructing the SOA-MHC, a comprehensive data validation process was implemented. The aim of the data validation was to assess the quality, accuracy, and consistency of the data gathered through various methods, such as stakeholder interviews, document reviews, observations, and benchmarking research. The principle of triangulation was applied to validate the collected data by comparing and cross-referencing information from multiple sources. This involved corroborating findings by comparing insights from different data collection methods, seeking convergence of evidence across various sources to strengthen the validity of the findings, and reconciling any discrepancies or contradictions through further investigation and clarification from relevant stakeholders.

Key stakeholders were engaged in the validation process to ensure that the collected data accurately reflected their perspectives and experiences. This involved sharing summarized findings and interpretations with stakeholders for their feedback and validation, soliciting input from subject matter experts to review the data and provide insights, and conducting validation workshops with a diverse group of stakeholders to discuss the data, gather additional feedback, and reach consensus on key findings and requirements.

A thorough assessment of the quality and completeness of the collected data was performed to ensure its suitability for constructing the SOA-MHC architecture. This involved reviewing the data for errors, inconsistencies, or missing information and applying data cleaning techniques to address any identified issues. Consistency checks ensured the coherence and logic of the data across different sources and data points, and an assessment of completeness identified gaps or areas requiring additional data collection.

An iterative approach to data validation was adopted, continuously refining the quality of the collected data throughout the research process. With progress through the data collection and analysis phases, the data were revisited and refined based on new insights. Maintaining an open feedback loop with stakeholders and subject matter experts allowed for ongoing input and validation, and detailed documentation of the data validation process, including the techniques applied, feedback received, and refinements made, provided transparency and traceability.

Employing these data validation techniques ensured that the data collected for constructing the SOA-MHC architecture was valid, reliable, and of high quality. The validated data served as a robust foundation for the design and development of the SOA-MHC architecture, which increased the credibility and trustworthiness of the research findings and recommendations.

### *3.3.3 SOA-MHC Design and Development*

The SOA-MHC proposed architecture follows a layered architecture design, comprising three main layers: Data Integration Layer, Service Layer, and Presentation Layer. The Data Integration Layer is responsible for extracting, transforming, and loading data from various sources into a centralized data repository. The Service Layer consists of a catalogue of reusable and interoperable services that expose data and analytics functionalities through APIs and microservices. The Presentation Layer provides a user-friendly interface for accessing and visualizing data insights.

The Data Integration Layer handles the ingestion and processing of data from diverse sources, such as EHRs. It employs Extract, Transform, Load (ETL) processes and data pipelines to cleanse, transform, and normalize the data before storing it in a staging area or data lake. The layer ensures data quality, consistency, and compatibility across different formats and structures.

The Service Layer is the core of the SOA-MHC design, encapsulating the business logic and data processing functionalities. It consists of a catalogue of loosely coupled and reusable services that can be easily composed and orchestrated to support various healthcare workflows. The services are designed following SOA principles, such as service contract, abstraction, and statelessness. The layer includes data services that expose data through APIs and microservices, as well as analytics services that provide real-time and batch processing capabilities.

The Presentation Layer focuses on delivering a user-friendly and intuitive interface for healthcare professionals to access and interact with the SOA-MHC. It includes the design of dashboards, reports, and data visualization tools that enable users to explore and derive insights from healthcare data. The layer incorporates role-based access control (RBAC) to ensure secure and personalized access to data and functionalities based on user roles and permissions.

Security and governance are critical aspects of the SOA-MHC. The architecture incorporates robust data security measures, such as encryption, access controls, and audit trails, to protect sensitive patient information and comply with healthcare regulations (e.g., HIPAA, GDPR). Data governance policies and procedures are established to ensure data quality, consistency, and ethical use throughout the data lifecycle.

Figure 2 illustrates the Service-Oriented Architecture for Big Data Management (SOA-MHC) tailored for a healthcare center. The architecture ensures strict adherence to security and governance standards, including HIPAA Compliance and Data Governance Policies, which are applied across all layers to protect sensitive patient information and maintain data integrity.

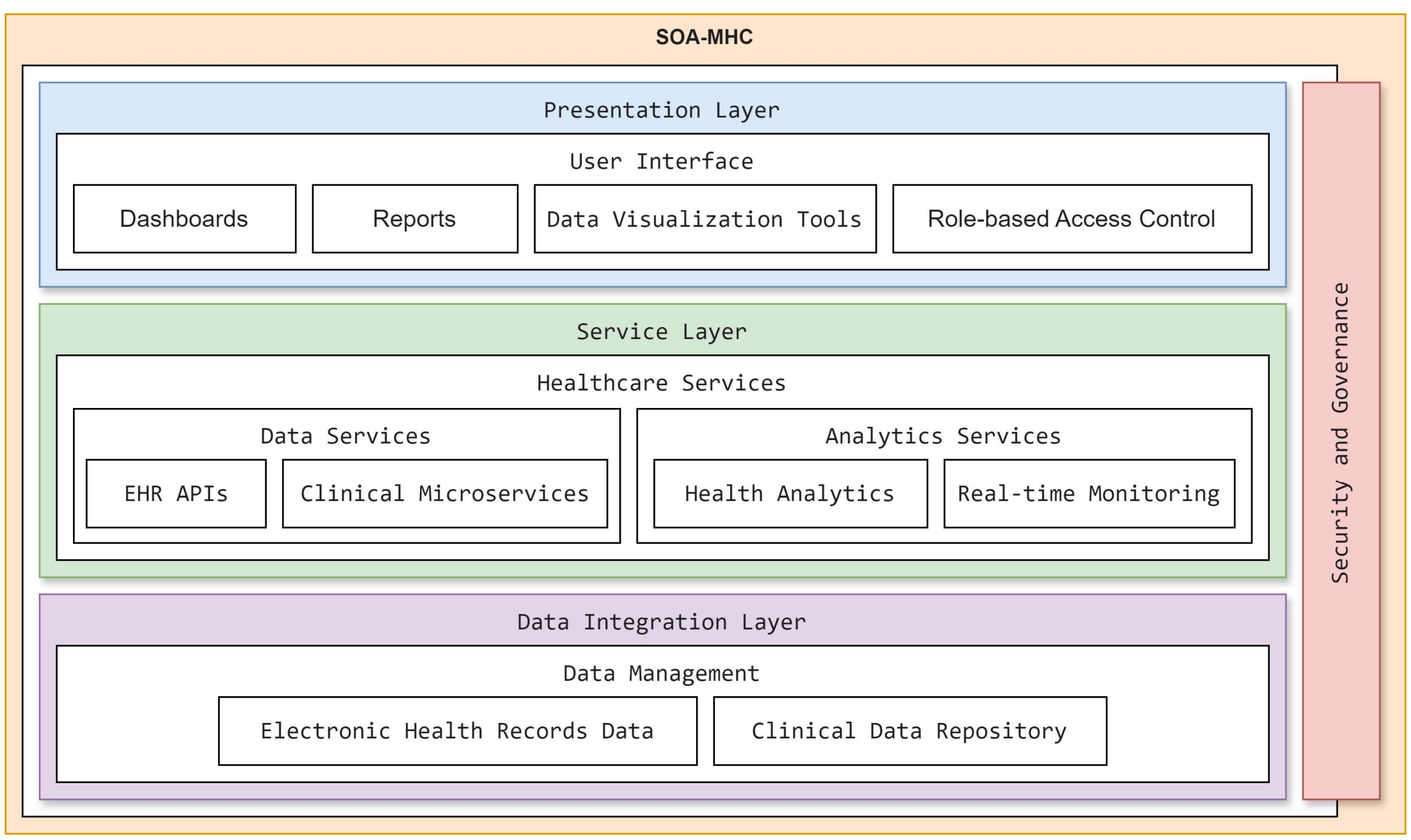

**Figure 2** SOA-MHC proposed architecture for a healthcare center.

*3.3.4 Implementation*

The Bahraini primary healthcare center will adopt the SOA-MHC architecture through a careful, phased approach, ensuring a smooth transition to the new system. Initially, infrastructure setup was carried out, involving the establishment of the necessary hardware and software to support the architecture. This setup included servers, databases, networking components, and the installation and configuration of various software tools and platforms. Subsequently, the center embarks on data migration, transferring existing healthcare data from numerous sources such as EHRs and legacy systems into a new Clinical Data Repository within the Data Integration Layer. The team developed scripts and employed Extract, Transform, Load (ETL) processes to facilitate this transfer, all that while maintaining a focus on data quality and consistency.

Service development followed, with services in the Service Layer being crafted in accordance with SOA principles and industry best practices. This stage saw the creation of APIs, microservices, and analytics components using suitable programming languages and frameworks. A key aim was to ensure these services were modular, reusable, and scalable to accommodate future developments.

For instance, in crafting the healthcare data service, a WSDL file (*healthcare_service.wsdl*) – Figure 3 will be created to describe the service's functions and the data involved. Additionally, a UDDI service registration file (*healthcare_service_uddi.xml*) will be developed to facilitate the service's discoverability within the UDDI registry.

The WSDL file describes the healthcare data service, including its operations, messages, and data types. It defines the *HealthcareDataService* with a single operation *getPatientData*. The *getPatientData* operation takes a *patient ID* as input and returns patient data as output. The PatientData complex type defines the structure of the patient data, including name, age, and medical history.

**WSDL file (*healthcare_service.wsdl*)**

```
<?xml version="1.0" encoding="UTF-8"?>
<definitions xmlns="http://schemas.xmlsoap.org/wsdl/"
        xmlns:soap="http://schemas.xmlsoap.org/wsdl/soap/"
        xmlns:tns="http://example.com/healthcareservice"
        targetNamespace="http://example.com/healthcareservice">

 <service name="HealthcareDataService">
  <port name="HealthcareDataPort" binding="tns:HealthcareDataBinding">
   <soap:address location="http://example.com/healthcareservice"/>
  </port>
 </service>

 <binding name="HealthcareDataBinding" type="tns:HealthcareDataPortType">
  <soap:binding style="document" transport="http://schemas.xmlsoap.org/soap/http"/>
```

**WSDL file (*healthcare_service.wsdl*)**

```
  <operation name="getPatientData">
   <soap:operation soapAction="http://example.com/healthcareservice/getPatientData"/>
   <input>
    <soap:body use="literal"/>
   </input>
   <output>
    <soap:body use="literal"/>
   </output>
  </operation>
 </binding>

 <portType name="HealthcareDataPortType">
  <operation name="getPatientData">
   <input message="tns:getPatientDataRequest"/>
   <output message="tns:getPatientDataResponse"/>
  </operation>
 </portType>

 <message name="getPatientDataRequest">
  <part name="patientId" type="xsd:string"/>
 </message>
 <message name="getPatientDataResponse">
  <part name="patientData" type="tns:PatientData"/>
 </message>

 <types>
  <xsd:schema targetNamespace="http://example.com/healthcareservice">
   <xsd:complexType name="PatientData">
    <xsd:sequence>
     <xsd:element name="name" type="xsd:string"/>
     <xsd:element name="age" type="xsd:int"/>
     <xsd:element name="medicalHistory" type="xsd:string"/>
    </xsd:sequence>
   </xsd:complexType>
  </xsd:schema>
 </types>

</definitions>
```

**Figure 3** The content of WSDL file.

The UDDI file – Figure 4 will be used to service registration for the healthcare data service. It provides information about the service, including its name, description, and access point. The UDDI registration allows the service to be discovered and consumed by other applications.

**UDDI service registration (healthcare_service_uddi.xml)**

```
<uddi:businessService xmlns:uddi="urn:uddi-org:api_v3">
 <uddi:name>Healthcare Data Service</uddi:name>
 <uddi:description>A service for accessing patient healthcare data</uddi:description>
 <uddi:bindingTemplates>
  <uddi:bindingTemplate>
   <uddi:accessPoint>http://example.com/healthcareservice</uddi:accessPoint>
   <uddi:tModelInstanceDetails>
    <uddi:tModelInstanceInfo tModelKey="uddi:example.com:healthcareservice"/>
   </uddi:tModelInstanceDetails>
  </uddi:bindingTemplate>
 </uddi:bindingTemplates>
</uddi:businessService>
```

**Figure 4** The content of UDDI file.

The integration and testing phase saw the newly developed services being combined with the Data Integration Layer and Presentation Layer to create a unified SOA-MHC system. Extensive testing will ensure the system's functionality, performance, and security are up to standard. This included different types of testing such as unit, integration, and user acceptance testing, which verified that the system met its requirements and performed as expected. For example, the system's SOAP communication will be confirmed during testing; a service consumer will send a SOAP request specifying a patient ID, and the service provider will return the appropriate patient data in a SOAP response. The file below in Figure 5 represents a SOAP request to the *getPatientData* operation. It includes the *patient ID* as the input parameter in the request body.

**SOAP request (get_patient_data_request.xml)**

```
<soapenv:Envelope xmlns:soapenv="http://schemas.xmlsoap.org/soap/envelope/"
          xmlns:tns="http://example.com/healthcareservice">
  <soapenv:Header/>
  <soapenv:Body>
   <tns:getPatientDataRequest>
    <patientId>123456</patientId>
   </tns:getPatientDataRequest>
  </soapenv:Body>
</soapenv:Envelope>
```

**Figure 5** The content of SOAP request file.

The file - Figure 6 below represents a SOAP response from the *getPatientData* operation. It includes the patient data (name, age, and medical history) in the response body.

**SOAP response (get_patient_data_response.xml)**

```
<soapenv:Envelope xmlns:soapenv="http://schemas.xmlsoap.org/soap/envelope/"
          xmlns:tns="http://example.com/healthcareservice">
  <soapenv:Header/>
  <soapenv:Body>
   <tns:getPatientDataResponse>
    <patientData>
     <name>Hasan Hameed Abdulla</name>
     <age>35</age>
     <medicalHistory>Hypertension, Diabetes</medicalHistory>
    </patientData>
   </tns:getPatientDataResponse>
  </soapenv:Body>
</soapenv:Envelope>
```

**Figure 6** The content of SOAP response file.

To safeguard sensitive patient data, security and governance measures will be put into place as stipulated by the SOA-MHC architecture. This included setting up access controls, encryption, and audit logs, alongside the establishment of data governance policies to guide stakeholders in ethical data management. User training and change management will also be integral to the implementation process. Healthcare professionals will receive training on using the system's features and managing data and change management techniques will be applied to encourage adoption and address resistance.

Before a full-scale launch, a pilot deployment will be conducted, allowing for real-world testing and feedback collection, which informed further refinements to the system. After a successful pilot, the system was rolled out across the center, with ongoing monitoring, support, and continuous improvement based on user feedback and performance metrics.

Figure 7 illustrates the flow of information and the interactions between the service provider, UDDI registry, and service consumer in the context of healthcare big data management using WSDL, SOAP, and UDDI.

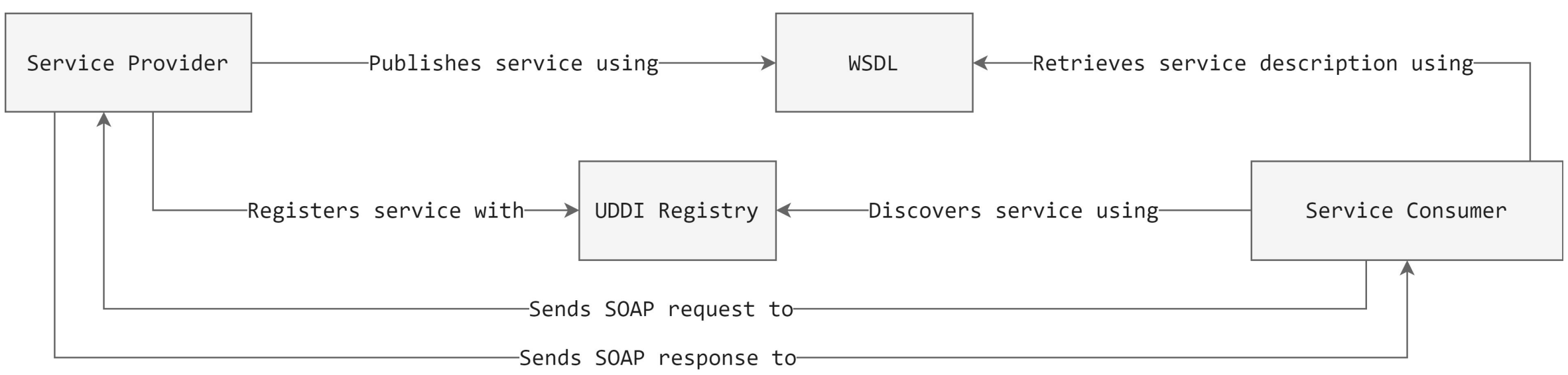

**Figure 7** WSDL, SOAP, and UDDI in healthcare big data management.

The SOA-MHC implementation will showcase the effective use of SOA principles in handling healthcare big data, with modular and reusable services enabling seamless system integration. The architecture's scalability and flexibility will lay a solid foundation for data management, providing healthcare professionals with essential information for better decision-making and patient care.

### 3.4 Demonstration

For the demonstration, the SOA-MHC was implemented in a scenario that closely resembled the authentic environment of the healthcare center. This ensured that the demonstration could provide relevant insights into the system's performance in an operational setting. The data and scenarios used were based on actual cases, tailored to exhibit the system's capacity to meet the center's needs.

The demonstration was executed by simulating a series of operations that are typically performed within the healthcare center. This included patient data management, data integration processes, and advanced data analytics. The demonstration showcased the artifact's interaction with existing systems and workflows, highlighting the benefits of the SOA-MHC in a real-world application.

During the demonstration, observations were meticulously recorded to document the artifact's performance and its interaction with users. This ensured that the demonstration could be later analyzed to identify successes, limitations, and areas where the SOA-MHC may require further refinement.

After the demonstration, the collected data were analyzed to assess the SOA-MHC's performance against the objectives outlined in the problem identification and objectives of a solution phases of DSRM. This analysis was crucial to determine whether the artifact met the requirements and to identify opportunities for improvement.

### 3.5 Evaluation

The evaluation phase, as delineated by the Design Science Research Methodology (DSRM), is essential for assessing the efficacy and utility of the developed artifact. In the case of the SOA-MHC, the evaluation step is designed to systematically measure how well the artifact addresses the solution provided to solve the problem within the Bahraini Primary Healthcare Centre.

In accordance with the Technology Acceptance Model, the evaluation phase commenced by setting forth clear and measurable objectives. These objectives were harmonized with the initial goals identified during the problem identification phase, as well as the stipulated requirements from the design and development stages. The objectives focused on evaluating key aspects such as system performance, user acceptance based on perceived ease of use and perceived usefulness, data accuracy, and the consequent impact on the efficacy of healthcare services delivery.

The Delphi process ensued through face-to-face meetings or video conferencing, with responses recorded and transcribed for subsequent analysis. The iterative nature of the Delphi technique permitted the refinement of ideas and consensus-building among participants.

Employing the Delphi technique, a meticulous evaluation methodology was formulated, encompassing both quantitative and qualitative assessments of the SOA-MHC. The results of the analysis were then meticulously synthesized to formulate a coherent set of findings that critically assessed the SOA-MHC's effectiveness. This synthesis delineated the extents to which the system either met, surpassed, or did not reach the predefined expectations, thereby casting light on the system's notable strengths and pinpointing areas ripe for enhancement.

Figure 8 illustrates the findings from the two rounds of the Delphi technique. Five criteria were evaluated on a scale of 0 to 5: Ease of Use, Usefulness, Attitudes, Behaviors, and Performance. The blue line represents the results for Round 1, while the orange line shows Round 2 results. Comparing the two rounds reveals an overall improvement from Round 1 to Round 2. Specifically, Ease of Use and Usefulness saw the most significant increases, with Round 2 outperforming Round 1. Smaller gains were observed for Attitudes. Behaviors scored low in both rounds, but a minor improvement occurred in Round 2. Performance was relatively high for both rounds, with a slight uptick in Round 2. The findings indicate enhanced outcomes across all criteria from Round 1 to Round 2, with the most notable progress in Ease of Use and Usefulness.

Based on the findings, recommendations for refining the SOA-MHC were developed. These recommendations aimed to address any deficiencies and to enhance the system's capabilities, ensuring that subsequent iterations would offer increased value to the healthcare center.

The entire evaluation process, including the methodology, data analysis, findings, and recommendations, was meticulously documented and reported. This documentation ensured transparency and provided a basis for further research and development.

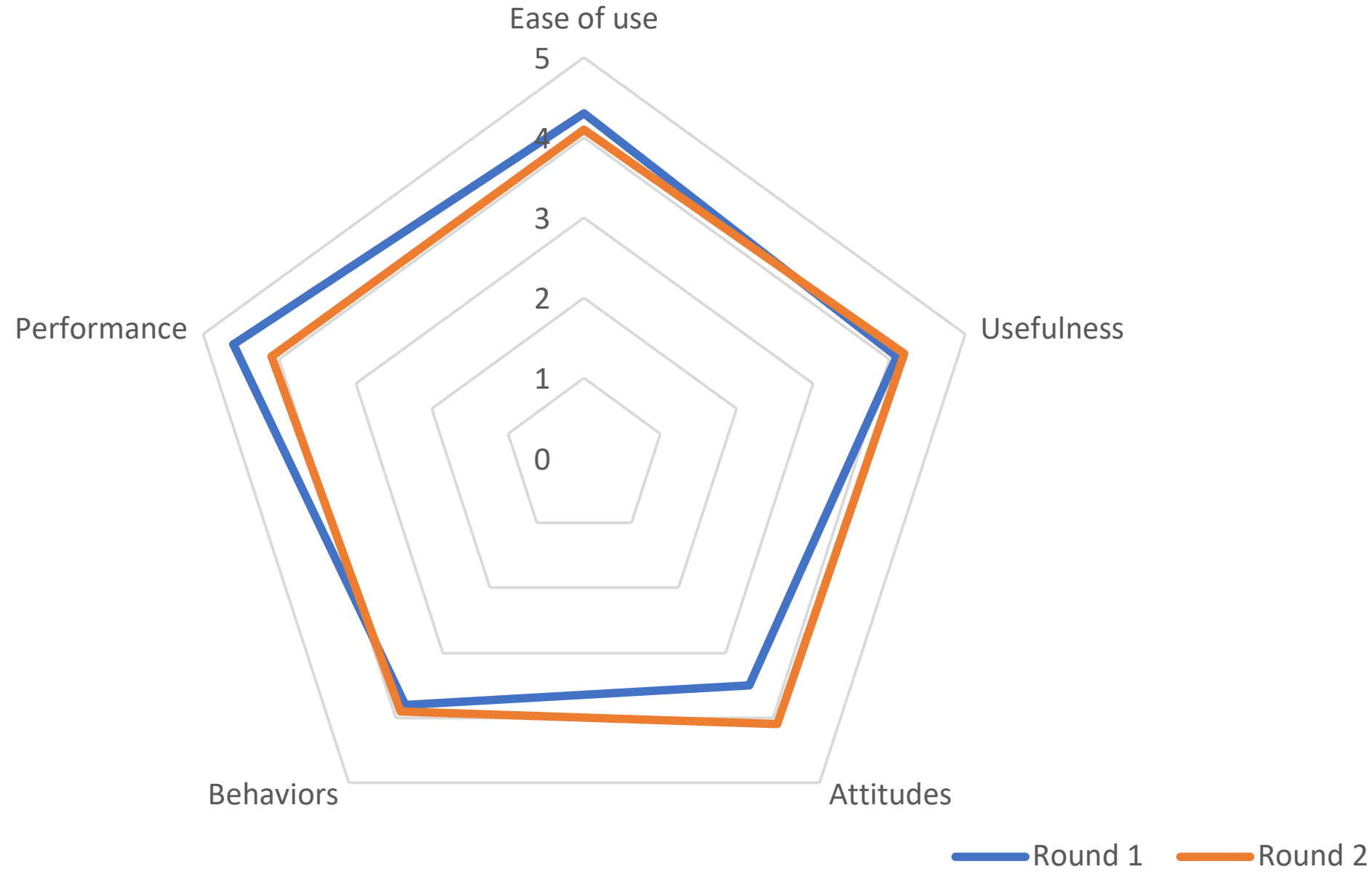

**Figure 8** Delphi technique results.

### 3.6 Communication

The act of communicating research through the development and publication of a research article represents a vital phase in the DSRM process. It encapsulates the entire journey of the SOA-MHC research project, from its inception to its thorough evaluation, and places the findings at the disposal of the scientific community. This formal presentation not only adds to the academic discourse but also stands as a point of reference for future innovations in healthcare technology.

## 4. Findings & Discussion

The development and implementation of the SOA-MHC yielded significant findings that shed light on the effectiveness of the proposed solution in addressing the challenges faced by the healthcare center. This section presents the key findings and discusses their implications for the healthcare Centre's data management practices and overall operational efficiency.

### 4.1 Improved Data Integration and Interoperability

One of the most notable findings of the SOA-MHC implementation was the significant improvement in data integration and interoperability across different systems and data sources within the healthcare center. The SOA-based architecture enabled seamless integration of data from electronic health records (EHRs), medical imaging systems, and other sources, creating a unified and comprehensive view of patient data. This integration eliminated data silos and facilitated the exchange of information between different departments and healthcare professionals, leading to enhanced collaboration and coordination of care.

The improved data integration and interoperability had several positive impacts on the healthcare center's operations. It reduced the time and effort required for healthcare professionals to access and retrieve relevant patient information, enabling faster and more informed decision-making. The availability of a comprehensive patient profile also facilitated the identification of potential health risks and the development of personalized treatment plans. Moreover, the standardization of data formats and protocols through the SOA-MHC promoted interoperability with external healthcare systems, enabling the smooth exchange of patient data with other healthcare providers and facilitating continuity of care.

### 4.2 Enhanced Data Analytics and Insights

The SOA-MHC empowered the healthcare center with advanced data analytics capabilities, enabling the extraction of valuable insights from the vast amount of collected data. The system's big data processing and analytics components allowed for real-time analysis of structured and unstructured data, uncovering patterns, trends, and correlations that were previously hidden.

The enhanced data analytics capabilities had a profound impact on various aspects of the healthcare center's operations. It enabled the identification of high-risk patients, allowing for proactive interventions and preventive care measures. The system's predictive analytics models helped forecast patient outcomes, resource utilization, and potential readmissions, enabling better planning and resource allocation. Furthermore, the insights derived from data analytics supported evidence-based decision-making, helping healthcare professionals make informed choices regarding treatment options, medication management, and care coordination.

The SOA-MHC also facilitated the generation of comprehensive reports and dashboards, providing healthcare administrators and managers with a clear overview of key performance indicators, patient outcomes, and resource utilization. These insights supported data-driven decision-making at the organizational level, enabling the identification of areas for improvement, optimization of processes, and effective resource allocation.

**4.3 Improved Operational Efficiency and Cost Savings**

The implementation of the SOA-MHC resulted in significant improvements in operational efficiency and cost savings for the healthcare center. The streamlined data management processes, automated data workflows, and real-time data availability reduced manual efforts and eliminated redundant tasks, freeing up healthcare professionals' time to focus on patient care.

The system's intelligent data processing and analytics capabilities enabled the identification of inefficiencies and bottlenecks in healthcare processes. By leveraging these insights, the healthcare center was able to optimize resource allocation, reduce waste, and improve overall productivity. The SOA-MHC also facilitated the implementation of lean healthcare practices, such as just-in-time inventory management and streamlined patient flow, resulting in reduced costs and improved operational efficiency.

Moreover, the SOA-MHC predictive analytics capabilities helped identify potential cost-saving opportunities, such as reducing unnecessary hospital readmissions, optimizing medication management, and minimizing the use of high-cost interventions. By proactively addressing these areas, the healthcare center was able to achieve significant cost savings without compromising the quality of care.

**4.4 Enhanced Patient Engagement and Satisfaction**

The SOA-MHC had a positive impact on patient engagement and satisfaction. The system's patient portal and mobile applications provided patients with secure access to their health information, enabling them to actively participate in their own care. Patients could view their medical records, lab results, medication lists, and upcoming appointments, fostering a sense of empowerment and ownership over their health.

The system's data analytics capabilities also enabled personalized patient communication and education. By analyzing patient data, the healthcare center could identify patients who would benefit from targeted health education materials, reminders for preventive screenings, and disease management programs. This personalized approach to patient engagement led to increased patient adherence to treatment plans, improved self-management of chronic conditions, and higher patient satisfaction scores.

Furthermore, the SOA-MHC integration with telemedicine and remote monitoring technologies enhanced patient access to care, particularly for those in remote or underserved areas. Patients could receive virtual consultations, remote monitoring of vital signs, and timely interventions, improving the continuity and quality of care while reducing the need for in-person visits.

**4.5 Challenges and Lessons Learned**

While the implementation of the SOA-MHC yielded numerous benefits, it also presented some challenges and valuable lessons learned. One of the main challenges was the initial resistance to change among healthcare professionals, who were accustomed to traditional data management practices. Overcoming this resistance required extensive training, change management efforts, and the demonstration of the system's tangible benefits to gain staff buy-in and adoption.

Another challenge was ensuring the security and privacy of patient data in the SOA-MHC. Implementing robust security measures, such as data encryption, access controls, and audit trails, was crucial to maintain the confidentiality and integrity of sensitive patient information. Regular security audits and compliance with relevant regulations, such as HIPAA and GDPR, were essential to mitigate potential data breaches and maintain patient trust.

The implementation process also highlighted the importance of effective data governance and data quality management. Establishing clear data governance policies, data standards, and data validation processes was crucial to ensure the accuracy, consistency, and reliability of the data used for decision-making. Regular data quality assessments and data cleansing efforts were necessary to maintain the integrity of the data and derive meaningful insights.

Lastly, the SOA-MHC implementation emphasized the significance of continuous improvement and stakeholder engagement. Regularly seeking feedback from healthcare professionals, patients, and other stakeholders helped identify areas for refinement and optimization. Incorporating user feedback and adapting the system to evolving needs and requirements ensured its long-term success and sustainability.

The findings and discussion of the SOA-MHC implementation highlight the significant benefits it brought to the Bahraini primary healthcare center. The improved data integration, enhanced analytics capabilities, operational efficiency gains, and patient engagement improvements demonstrate the effectiveness of the SOA-based approach in addressing the challenges of big data management in healthcare. While challenges were encountered during the implementation process, the lessons learned serve as valuable insights for future implementations and the continuous improvement of the system. The SOA-MHC success in transforming the healthcare center's data management practices and improving patient care delivery serves as a model for other healthcare organizations seeking to leverage the power of big data and service-oriented architectures in the healthcare domain.

## 5. Conclusion

The implementation of the SOA-MHC has demonstrated the transformative potential of leveraging advanced technologies and architectural principles to address the challenges of big data management in healthcare. This research aimed to develop a scalable, flexible, and secure solution that could effectively integrate and analyze healthcare data from diverse sources, empowering healthcare professionals to make data-driven decisions and improve patient care delivery.

The findings of this study highlight the significant benefits achieved through the SOA-MHC. The improved data integration and interoperability enabled by the SOA-based architecture eliminated data silos and facilitated seamless information exchange among different systems and departments within the healthcare center. This integration provided healthcare professionals with a comprehensive view of patient data, leading to enhanced collaboration, faster decision-making, and more coordinated care delivery.

Moreover, the SOA-MHC system's advanced data analytics capabilities empowered the healthcare center to extract valuable insights from the vast amount of collected data. The real-time analysis of structured and unstructured data uncovered patterns, trends, and correlations that supported evidence-based decision-making and personalized patient care. The system's predictive analytics models enabled the identification of high-risk patients, optimization of resource allocation, and proactive interventions, ultimately improving patient outcomes and operational efficiency.

Implementing the SOA-MHC also resulted in significant operational efficiency gains and cost savings for the healthcare center. The streamlined data management processes, automated workflows, and real-time data availability reduced manual efforts, eliminated redundancies, and optimized resource utilization. The system's intelligent data processing and analytics capabilities facilitated the identification of inefficiencies, enabling the healthcare center to implement lean practices and achieve cost savings without compromising the quality of care.

Furthermore, the SOA-MHC had a positive impact on patient engagement and satisfaction. The patient portal and mobile applications provided patients with secure access to their health information, fostering a sense of empowerment and active participation in their care. The personalized patient communication and education enabled by data analytics led to increased patient adherence, improved self-management of chronic conditions, and higher patient satisfaction scores.

However, implementing the SOA-MHC also presented challenges and valuable lessons learned. Overcoming initial resistance to change among healthcare professionals required extensive training and change management efforts. Ensuring the security and privacy of patient data necessitated implementing robust security measures and compliance with relevant regulations. Effective data governance and quality management were crucial to maintaining the accuracy, consistency, and reliability of the data used for decision-making. Continuous improvement and stakeholder engagement were essential for refining the system and adapting to evolving needs and requirements.

The successful implementation of the SOA-MHC in the Bahraini primary healthcare center serves as a model for other healthcare organizations seeking to leverage big data and service-oriented architectures to transform their data management practices and improve patient care delivery. This study's findings contribute to the growing body of knowledge on the application of SOA principles in healthcare and provide valuable insights for practitioners and researchers in the field.

For future work, more research can be done to investigate how the SOA-MHC affects patient results, cost-efficiency, and health management at a population level. Additionally, future studies can investigate the scalability and applicability of the SOA-MHC model in different healthcare settings, such as specialized clinics, hospitals, and nationwide health systems. As healthcare data grows in volume, variety, and velocity, adopting service-oriented architectures and big data management solutions will become increasingly crucial for healthcare organizations to harness the full potential of their data assets and deliver high-quality, patient-centered care.

In conclusion, the development and implementation of the SOA-MHC in the Bahraini primary healthcare center have demonstrated the transformative potential of leveraging advanced technologies and architectural principles to address the challenges of big data management in healthcare. The findings of this study highlight the significant benefits achieved in terms of improved data integration, enhanced analytics capabilities, operational efficiency gains, and patient engagement improvements. While challenges were encountered during the implementation process, the lessons learned serve as valuable insights for future implementations and the continuous improvement of the system. The success of the SOA-MHC in transforming the healthcare center's data management practices and improving patient care delivery underscores the importance of adopting innovative approaches to harness the power of big data in healthcare. As healthcare organizations continue to navigate the complexities of the big data landscape, the principles and practices exemplified by the SOA-MHC provide a roadmap for leveraging data-driven insights to enhance healthcare delivery and ultimately improve patient outcomes.

## References

[1] M. Fugini, J. Finocchi and P. Locatelli, "A big data analytics architecture for smart cities and smart companies," *Big Data Res.,* vol. 24, p. 100192, May 2021.

[2] S. Kumar and M. Singh, "Big data analytics for healthcare industry: impact, applications, and tools," *Big Data Min. Anal.,* vol. 2, p. 48–57, March 2019.

[3] F. Li, X. Yu, R. Ge, Y. Wang, Y. Cui and H. Zhou, "BCSE: Blockchain-based trusted service evaluation model over big data," *Big Data Min. Anal.,* vol. 5, p. 1–14, March 2022.

[4] M. Volk, D. Staegemann, A. Saxena, J. Hintsch, N. Jamous and K. Turowski, "Lowering Big Data Project Barriers: Identifying System Architecture Templates for Standard Use Cases in Big Data," in *Proceedings of the 19th International Conference on Smart Business Technologies*, 2022.

[5] J. Saltz and S. Lahiri, "The need for an enterprise risk management framework for big data science projects," in *Proceedings of the 9th International Conference on Data Science, Technology and Applications*, Lieusaint - Paris, France, 2020.

[6] E. A. Bayrak and P. Kirci, "A brief survey on big data in healthcare," *Int. J. Big Data Anal. Healthc.,* vol. 5, p. 1–18, January 2020.

[7] N. Mutiah and F. Febriyanto, "Perancangan Arsitektur Enterprise FMIPA UNTAN Menggunakan Kerangka Kerja TOGAF Berbasis SOA," *J. SIST. INF. BISNIS,* vol. 12, p. 116–123, December 2022.

[8] J. E. Kasten, "Big Data Applications in Vaccinology," *Int. J. Big Data Anal. Healthc.,* vol. 6, p. 59–80, April 2021.

[9] J. N. Undavia and A. M. Patel, "An article on big data analytics in healthcare applications and challenges," *Int. J. Big Data Anal. Healthc.,* vol. 5, p. 58–64, July 2020.

[10] C. Dias, M. Santos and F. Portela, "A SWOT analysis of big data in healthcare," in *Proceedings of the 6th International Conference on Information and Communication Technologies for Ageing Well and e-Health*, Prague, 2020.

[11] D. Roosan, M. Karim, J. Chok and M. Roosan, "Operationalizing healthcare big data in the electronic health records using a heatmap visualization technique," in *Proceedings of the 13th International Joint Conference on Biomedical Engineering Systems and Technologies*, Valletta, 2020.

[12] L. Ilie, M. A. Moisescu, S. I. Caramihai and J. Culita, "Enterprise architecture role in hospital management systems development," in *2021 23rd International Conference on Control Systems and Computer Science (CSCS)*, Bucharest, 2021.

[13] E. Heiland, P. Hillmann and A. Karcher, "Enterprise architecture model transformation engine," in *Proceedings of the 10th International Conference on Operations Research and Enterprise Systems*, Online Streaming, — Select a Country —, 2021.

[14] M. M. Nezhad and F. S. Aliee, "Improving enterprise architecture mining methods," in *2023 7th Iranian Conference on Advances in Enterprise Architecture (ICAEA)*, Tehran, Iran, Islamic Republic of, 2023.

[15] M. Paiva, A. Vasconcelos and B. Fragoso, "Using enterprise architecture to model a reference architecture for industry 4.0," in *Proceedings of the 22nd International Conference on Enterprise Information Systems*, Prague, 2020.

[16] M. G. Mariam and B. Bygstad, "What can enterprise architecture do for healthcare A framework of antecedents and benefits," *Electron. Gov. Int. J.,* vol. 15, p. 213, 2019.

[17] M. Ribeiro and A. Vasconcelos, "MedBlock: Using blockchain in health healthcare application based on blockchain and smart contracts," in *Proceedings of the 22nd International Conference on Enterprise Information Systems*, Prague, 2020.

[18] J. G. G. N. B. S. M. A. &. L. S. T. Jonnagaddala, "Adoption of enterprise architecture for healthcare in AeHIN member countries.," *BMJ health & care informatics,* vol. 1, p. 27, 2020.

[19] C. Apsey, M. Jawad, M. Daschel, D. Woosey, C. Limb and M. Wilde, "Does thrombolysis for pulmonary embolism reduce the risk of chronic complications?," *Acute Med.,* vol. 20, p. 15–17, 2021.

[20] M. Wang, . S. Li, T. Zheng, N. Li, Q. Shi, X. Zhuo, R. Ding and Y. Huang , "Big Data Health Care Platform With Multisource Heterogeneous Data Integration and Massive High-Dimensional Data Governance for Large Hospitals: Design, Development, and Application," *JMIR Medical Informatics,* vol. 10, no. 4, 2022.

[21] K. Osei-Tutu, S. Hasavari and Y.-T. Song, "Blockchain-based enterprise architecture for comprehensive healthcare information exchange (HIE) data management," in *2020 International Conference on Computational Science and Computational Intelligence (CSCI)*, Las Vegas, NV, USA, 2020.

[22] W. Daoudi, K. Doumi and L. Kjiri, "Complexity and adaptive enterprise architecture," in *Proceedings of the 23rd International Conference on Enterprise Information Systems*, Online Streaming, — Select a Country —, 2021.

[23] L. Jing and O. Manta, "Research on Digital Campus Construction Based on SOA Service Architecture," in *Proceedings of the 2022 3rd International Symposium on Big Data and Artificial Intelligence*, New York, NY, USA, 2022.

[24] Z. Li and E. J. Pino, "D&D: A distributed and disposable approach to privacy preserving data analytics in user-centric healthcare," in *2019 IEEE 12th Conference on Service-Oriented Computing and Applications (SOCA)*, Kaohsiung, 2019.

[25] P. W. Handayani, A. A. Pinem, Q. Munajat, F. Azzahro, A. N. Hidayanto, D. Ayuningtyas and A. Sartono, "Health referral enterprise architecture design in Indonesia," *Healthc. Inform. Res.,* vol. 25, p. 3–11, January 2019.

[26] M. Coelho, A. Vasconcelos and P. Sousa, "Privacy by design enterprise architecture patterns," in *Proceedings of the 23rd International Conference on Enterprise Information Systems*, Online Streaming, — Select a Country —, 2021.

[27] C. Teixeira, A. Vasconcelos, P. Sousa and M. Marques, "Enterprise Architecture Patterns for GDPR Compliance," in *Proceedings of the 23rd International Conference on Enterprise Information Systems*, Online Streaming, — Select a Country —, 2021.

[28] J. Patel, "An effective and scalable data modeling for enterprise big data platform," in *2019 IEEE International Conference on Big Data (Big Data)*, Los Angeles, CA, USA, 2019.

[29] O. Rajaei, S. R. Khayami and M. S. Rezaei, "Smart Hospital Technologies Investigation," in *2023 7th Iranian Conference on Advances in Enterprise Architecture (ICAEA)*, Tehran, Iran, Islamic Republic of, 2023.

[30] S. Raza, "A machine learning model for predicting, diagnosing, and mitigating health disparities in hospital readmission," *Healthcare Analytics,* vol. 2, p. 100100, November 2022.

[31] Y. Masuda, A. Zimmermann, D. S. Shepard, R. Schmidt and S. Shirasaka, "An adaptive enterprise architecture design for a digital healthcare platform : Toward digitized society – industry 4.0, society 5.0," in *2021 IEEE 25th International Enterprise Distributed Object Computing Workshop (EDOCW)*, Gold, 2021.

[32] S. Petersen and T. Evjen, "Enterprise architecture to identify the benefits of enterprise building information model data: An example from healthcare operations," in *Proceedings of the 24th International Conference on Enterprise Information Systems*, Online Streaming, — Select a Country —, 2022.

[33] S. Guraya, "The evolving landscape of digital health using big data analytics for personalized healthcare," *Adv. Biomed. Health Sci.,* vol. 1, p. 5, 2022.

[34] W. Noonpakdee, "The analysis of big data architecture for healthcare service, a case study of a public hospital," *Int. J. Manag. Knowl. Learn.,* vol. 11, p. 1–8, 2022.

[35] P. K. D. Pramanik, S. Pal and M. Mukhopadhyay, "Big data and big data analytics for improved healthcare service and management," *Int. J. Priv. Health Inf. Manag.,* vol. 8, p. 13–51, January 2020.

[36] H. Istanto, M. D. Arza, D. R. Irkhamna, D. F. Murad and S. Maryam, "Omnichannel architecture enterprise modeling on the core processes of the culinary industry using the TOGAF architecture development method (ADM)," in *2022 International Conference on Electrical Engineering and Informatics (ICELTICs)*, Banda, 2022.

[37] S. M. Zanjani, S. M. Zanjani, H. Shahinzadeh, M. Moazzami, F. Ebrahimi and N. Nourmahnad, "Leveraging big data intelligence for electronic health data collection and mining in smart innovative cities: Approaches and applications," in *2023 7th Iranian Conference on Advances in Enterprise Architecture (ICAEA)*, Tehran, Iran, Islamic Republic of, 2023.

[38] M. J. Lange M, "An experts' perspective on enterprise architecture goals, framework adoption and benefit assessment.," in *Enterprise Distributed Object Computing Conference Workshops (EDOCW). 2011 15th IEEE International*, 2011.

[39] M. N. H. D. M. A. &. Y. D. Alwi, "Efficiency and effectiveness: enterprise architecture strategies for healthcare service.," *International Journal Software Engineering and Computer Science (IJSECS),* vol. 3, no. 386-397, p. 3, 2023.

[40] A. Qostal, A. Moumen and Y. Lakhrissi, "Big data, Hadoop and spark for employability: Proposal architecture," in *Proceedings of the 2nd International Conference on Big Data, Modelling and Machine Learning*, Kenitra, 2021.

[41] L. Moraes, P. Jardim and C. Aguiar, "Design principles and a software reference architecture for big data question answering systems," in *Proceedings of the 25th International Conference on Enterprise Information Systems*, Prague, 2023.

[42] Y. Wang, "Design and construction of enterprise financial data sharing service center based on big data technology," in *Proceedings of the 2nd International Conference on Big Data Economy and Digital Management, BDEDM 2023, January 6-8, 2023, Changsha, China*, Changsha, People's Republic of China, 2023.

[43] J. P. C. Castro, L. M. F. Romero, A. C. Carniel and C. D. Aguiar, "FAIR principles and big data: A software reference architecture for open science," in *Proceedings of the 24th International Conference on Enterprise Information Systems*, Online Streaming, — Select a Country —, 2022.

[44] X. Liu, X. Wang, S. Zhuo and F. Zhang, "The impact of COVID-19 on nurses' mental healthcare and sustainability by data analysis," in *Proceedings of the 1st International Conference on Health Big Data and Intelligent Healthcare*, Sanya, 2022.

[45] J. N. Undavia and A. M. Patel, "Big Data Analytics in Healthcare," *Int. J. Big Data Anal. Healthc.,* vol. 5, p. 19–27, January 2020.

[46] D. Huth, M. Vilser, G. Bondel and F. Matthes, "Empirical task analysis of data protection management and its collaboration with enterprise architecture management," in *Proceedings of the 22nd International Conference on Enterprise Information Systems*, Prague, 2020.

[47] A. Cuzzocrea and S. Soufargi, "QFLS: A cloud-based framework for supporting big healthcare data management and analytics from big data lakes: Definitions, requirements, models and techniques," in *Proceedings of the 12th International Conference on Data Science, Technology and Applications*, Rome, 2023.

[48] F. Mohammed and M. Naaz, "Big Data Analytics: Challenges and Applications in Healthcare," *Int. J. Sci. Res. (Raipur),* vol. 12, p. 834–838, September 2023.

[49] J. N. Undavia and A. M. Patel, "Big Data Analytics in Healthcare," *Int. J. Big Data Anal. Healthc.,* vol. 5, p. 19–27, January 2020.

[50] S. Kumar and M. Singh, "Big data analytics for healthcare industry: impact, applications, and tools," *Big Data Min. Anal.,* vol. 2, p. 48–57, March 2019.

[51] Y. Li, Y. Bo, Y. Zhao and X. Jiao, "Smart elderly care system and health management empowered by big data: A case study of the innovation of the intelligent wheelchair application," in *Proceedings of the 1st International Conference on Health Big Data and Intelligent Healthcare*, Sanya, 2022.

[52] Y. P. Putra and A. Hadiana, "Designing enterprise architecture for public health center based on TOGAF architecture development method," *IOP Conf. Ser. Mater. Sci. Eng.,* vol. 879, p. 012163, July 2020.

[53] K. Peffers, M. Rothenberger, T. Tuunanen and R. Vaezi, "Design Science Research Evaluation," in *Design Science Research in Information Systems. Advances in Theory and Practice*, Springer Berlin Heidelberg, 2012, p. 398–410.

[54] A. Hevner and S. Chatterjee, "Design Science Research in Information Systems," in *Design Research in Information Systems*, Springer US, 2010, p. 9–22.

[55] J. vom Brocke, A. Hevner and A. Maedche, "Introduction to Design Science Research," in *Design Science Research. Cases*, Springer International Publishing, 2020, p. 1–13.

[56] P. Offermann, O. Levina, M. Schönherr and U. Bub, "Outline of a design science research process," in *Proceedings of the 4th International Conference on Design Science Research in Information Systems and Technology - DESRIST '09*, 2009.

[57] K. Ndlovu, R. E. Scott and M. Mars, "Interoperability opportunities and challenges in linking mhealth applications and eRecord systems: Botswana as an exemplar," *BMC medical informatics and decision making,* vol. 21, no. 1, 2021.

[58] K. Adnan, R. Akbar, S. W. Khor and A. Bin Amanat Ali , "Role and Challenges of Unstructured Big Data in Healthcare," in *Data Management, Analytics and Innovation. Advances in Intelligent Systems and Computing*, Singapore, 2020.

[59] V. Palanisamy and R. Thirunavukarasu, "Implications of big data analytics in developing healthcare frameworks – A review," *Journal of King Saud University - Computer and Information Sciences,* vol. 31, no. 4, pp. 415-425, 2019.

## **Appendix-1**- The Case study Protocol Document including data collection templates (1,2) *

| Sec # | Topic | Content |
|---|---|---|
| **1** | Overview | This research aims to benchmark the technological applications'data of the selected **Bahraini E-01** in the **P**rimary Healthcare Centers. |
| **2** | Research Questions-1 | **Q1. What are the commonly used functions in the Bahraini E-01?**<br>Q1.1 What are the Business Strategic objectives of **E-01**?<br>Q1.2 What are the Business Units& Actors/Roles of **E-01**?<br>Q1.3 What are the Business Functions of **E-01**?<br>Q1.4 What are the Business Services of **E-01**?<br>Q1.5 What are the Business Processes of **E-01**? |
| **3** | Research Questions-2 | **Q2. What are the ICT Applications supporting the Bahraini E-01?**<br>Q2.1 What are the ICT Applications supporting the Units &Functions?<br>Q2.2 What are the ICT Applications supporting the Processes?<br>Q2.3 What are the ICT Applications supporting the Services?<br>Q2.4 How to generate a visual pattern to express the relationships?<br>Q2.5 What are the gaps in the ICT Application portfolio? |
| **4** | Data collection matrix |  |
| **5** | Data collection matrix |  |

| Q# | Research Questions | Evidence (Tools) | Rationale | Data collection |
|---|---|---|---|---|
| **Q1** | What are the banks business functions? | | Develop BA of the E-01 | **Primary Sources:**<br>Interviews<br><br>**Secondary Sources:**<br>LR, Documentary, Web |
| 1.2 | How to examine the bus and tech? | Template I, II | Tailor the architecture definition document | |
| 1.3 | What are the BSO? | Template II | Align objectives with EA objectives. | |
| 1.4 | What are the BU & BA/R? | Template II | Define the current BA | |
| 1.5 | What are the BS? | Template II | Define the current business architecture. | |
| 1.6 | What are the BF? | Template II | Define the current business architecture. | |
| 1.7 | What are the BP? | Template II | Define the current business architecture. | |

| Q# | Research Questions | Evidence (Tools) | Rationale | Data collection technique |
|---|---|---|---|---|
| Q2 | What are the banks technological application? | | | |
| 2.1 | What AppN w.r.t. BU, BA/R, BF? | Template II | Traceability | **Primary Sources:**<br>Interviews<br>**Secondary Sources:**<br>LR, Documentary, Web |
| 2.2 | What AppN, AppS, AppF w.r.t BS? | Template II | Traceability | |
| 2.3 | What AppN, AppP w.r.t. BP? | Template II | Traceability | |
| 2.4 | What are the gaps in AppN? | Template II | Benchmark the findings | |

| | | |
|---|---|---|
| 6 | Template I | Business architecture (BA) |

**Business architecture (BA)**

A. Business Strategic Objectives

| Business Objectives ID | Business Objectives | IT objectives ID | IT bjectives |
|---|---|---|---|
| | | | |

B. Business Units and Actors/Roles

| Unit ID | Unit Name | Unit Parent | Unit Description | Actor/Role ID | Actor/Role |
|---|---|---|---|---|---|
| | | | | | |

C. Business Functions

| Unit ID | Function ID | Function Name | Function Description | Function Classification | Strategic Objective(s) ID |
|---|---|---|---|---|---|
| | | | | | |

D. Business Services

| Service-ID | Service Name | Service Description | Function- ID |
|---|---|---|---|
| | | | |

E. Business Processes

| Proc- ID | Proc- Name | Proc Description | Proc Class | Proc I/P | Supplier of input | Process Outputs | Consumer of Output |
|---|---|---|---|---|---|---|---|
| | | | | | | | |

| | | |
|---|---|---|
| 7 | Template II | Information systems architecture (ISA) |

**Information systems architecture (ISA)**

A. Baseline Applications

| Application ID | Application Name | Application Version | Application Description | Vendor |
|---|---|---|---|---|
| | | | | |

B. Application Portfolio Summary

| Application ID | Unit ID | Function ID | App Name | App Description | Business Owner | Application Status |
|---|---|---|---|---|---|---|
| | | | | | | |

C. Application Portfolio Details

| Business Function ID | Process ID | App. ID | Functional Component | Primary Users | Application Type |
|---|---|---|---|---|---|
| | | | | | |

| | | |
|---|---|---|
| 7 | Template II | Technology/Infrastructure architecture (TA) |

**Technology/Infrastructure architecture (TA)**

A. Baseline Technology components

| Tech ID | Tech Name | Tech Service | Technology Function | Technology Function Vendor |
|---|---|---|---|---|
| | | | | |

* Adwan, E.J. and Al-Soufi, A., 2020. Enterprise Architecture based ISA model development for ICT benchmarking in Construction-Case Study. Int J Inf & Commun Technol, 8(2).